\documentclass[preprint2]{aastex}

\def\lesssim{\mathrel{\hbox{\rlap{\hbox{\lower4pt\hbox{$\sim$}}}\hbox{$<$}}}}
\def\gtrsim{\mathrel{\hbox{\rlap{\hbox{\lower4pt\hbox{$\sim$}}}\hbox{$>$}}}}

\begin{document}

\title{Morphology-Density Relation for Simulated Clusters of Galaxies in
Cold Dark Matter Dominated Universes}


\author{Takashi Okamoto}
\affil{Div. of Physics, Grad. School of Science, Hokkaido University, Sapporo 060-0810, Japan}
\email{okamoto@astro1.sci.hokudai.ac.jp}

\and

\author{Masahiro Nagashima}
\affil{National Astronomical Observatory, Mitaka, Tokyo 181-8588, Japan}
\email{masa@th.nao.ac.jp}

\begin{abstract}
We present a model to investigate the formation and evolution
of cluster galaxies using cosmological high-resolution $N$-body
simulations.
The $N$-body simulations are used to construct merging history
trees of dark halos.
Gas cooling, star formation, supernova feedback and mergers
of galaxies within dark halos are included by using simple prescriptions
taken from semi-analytic models of galaxy formation.
In this paper, we examine the merger-driven bulge formation model and
represent the morphology-density relation of cluster galaxies at $z = 0$.
We find that this morphological evolution model can explain the
distribution of elliptical galaxies in the clusters well and
cannot reproduce the distribution of S0 galaxies.
This result suggests that the elliptical galaxies are mainly formed by the
major mergers, while, in the S0 formation, the processes other than major
mergers play important role.
\end{abstract}



\section{INTRODUCTION}

It is well established that galaxy populations vary with
the density of neighbouring galaxies in clusters of galaxies
\citep{dre80}.
This morphology-density relation (hereafter MDR) indicates that dynamical
processes that depend on environment of each galaxy mainly affect the final
configuration of the stellar component.
This impression seems to be supported by the Hubble Space Telescope images
of clusters at intermediate redshifts,
which show an abnormally high proportion of spiral and irregular
types at $z \sim 0.5$ and increase of S0 fraction toward the present time
\citep{cou94, dre97}.

Some mechanisms that may transform one morphological type into another
have been proposed, for example, ram-pressure stripping of interstellar medium
of spirals by a
intracluster medium \citep{gun72, fn99}, galaxy harassment by cumulative tidal
interactions in clusters (Moore et al. 1996, 1998),
and galaxy mergers  \citep{too77}.
$N$-body simulations have confirmed that merging disk galaxies
produce galaxies resembling ellipticals as merger remnants
(e.g. Barnes 1996),
so that the galaxy merger is a favorite explanation for the predominance
of ellipticals and S0s in rich clusters
\citep{kau93, kau96, bau96}.

The purpose of our study is to check this {\it merger hypotheses} by
investigating the MDR for the simulated clusters in the cosmological context.
For this purpose, it is difficult in the present situation to use numerical
simulations including both gravity and gas dynamics
(e.g. Katz et al. 1992;  Evrard et al. 1994),
because such simulations are expensive in CPU time and so only a
limited range of parameter space can be explored with an insufficient spatial
resolution.

Semi-analytic modeling of galaxy formation has already proved to be a powerful
technique (e.g. Kauffmann et al. 1993; Cole et al. 1994).
However, we cannot identify the position of galaxies by such approaches,
because these models follow the collapse and merging
histories of dark halos by using a probabilistic method on the mass
distribution based upon an extension  of the Press-Schechter
formalism \citep{pre72, bon91, bow91, lac93}.

One approach to identify the position and velocity of each galaxy is to
track the merging histories of dark halos by using $N$-body simulations
and to combine them with the simple prescriptions of the semi-analytic
models \cite{rou97, kau99, ben00}.
Their schemes, however, do not deal with the substructures within dark halos.
Since the dynamics within clusters may strongly affects the evolution of
cluster galaxies (Okamoto \& Habe 1999, 2000), we should use a new galaxy
tracing method.

In this paper, we adopt the galaxy tracing method provided
by Okamoto \& Habe (1999, 2000), which enables us to trace the individual
galactic dark halos within dense environments using high-resolution $N$-body
simulations, that is, we can obtain
the three-dimensional distribution of the galaxies within clusters.
Our formula of modeling gas cooling, star formation, supernova feedback,
and galaxy mergers is directly taken from the previous semi-analytic work.
We also adopt a merger-driven scheme for the production of galactic bulges
and the way of the morphological classification based on bulge-to-disk
ratios as earlier studies \cite{kau93, bau96}.

In \S 2, we describe a brief outline of the galaxy formation model used here.
We show the results about the morphology of the galaxies
in \S 3. These results are discussed in \S 4.

\section{MODEL}
We exam the evolution of cluster galaxies in
the standard cold dark matter (SCDM) universe
($\Omega_0 = 1, h \equiv H_0/100$km s$^{-1}$ Mpc$^{-1}=0.5, \sigma_8 = 0.67$)
and the open CDM (OCDM) universe
($\Omega_0 = 0.3, h = 0.7, \sigma_8 = 1.0$).
The baryon density is set to $\Omega_b = 0.1$ and $0.03$
for SCDM and OCDM, respectively.

The outline of the procedures of galaxy formation is as follows.
At first, the merging path of galactic halos are realized by
the cosmological high-resolution $N$-body simulations.
Next, in each merging path, evolution of the baryonic component,
namely, gas cooling, star formation, and supernova feedback,
are calculated based on Kauffmann et al. (1993) and Cole et al. (1994).
We refer a system consisting of the stars and cooled gas as a galaxy.
When two or more dark halos merge together, we estimate the merging
time-scale based on the dynamical friction time-scale.
When the merging of galaxies occurs, we change the morphology of the
merger remnant by the type of the merger.
Finally, we calculate the luminosity and color of each galaxy.
Through the above procedures, we obtain the morphological distribution
of cluster galaxies.

\subsection{$N$-Body Calculation and Merging of Dark Halos}
The merging histories of galactic dark halos are realized
using the same method and simulation data by Okamoto \& Habe (2000).
Note that they trace {\it halo-stripped galaxies} as well as the
galactic halos because halo disruption is probably due to
lack of dissipative processes which are not included in the
$N$-body simulations \citep{sum95}.
The merging histories are constructed with a 0.5 Gyr
time step in their paper, while here we adopt half of the time step
for high redshifts ($z \gtrsim 2.0$) at which the galactic halos form
and merge violently \citep{oka00}. This improvement, however, hardly changes
our results.

\subsection{Model of Galaxy Formation}
The following prescriptions are almost the same as the previous ordinary
semi-analytic models.

For simplicity, a dark halo is modeled as an isothermal sphere
whose mass and radius are taken from $N$-body data.
The source of the diffuse gas in a halo is hot gas contained in
its progenitor halos and in the accreting matter.
The baryon fraction of the accreting matter is defined as
$f_{\rm b} = \frac{\Omega_{\rm b}}{\Omega_0}$.
We assume that the hot gas has the distribution that parallels to that of the
dark matter with the virial temperature of the halo.
When a galactic halo is tidally stripped, the hot gas in the halo is also
stripped proportional to the amount of the stripped dark matter.

The cooling time-scale, $\tau_{\rm cool}$, is obtained as a function of the
radius from the hot gas density profile, the temperature of the gas,
and the cooling function $\Lambda(T)$ as follows,
\begin{equation}
\tau_{\rm cool} =
\frac{3}{2}\frac{\rho(r)}{\mu m_{\rm p}}\frac{kT}{n_{\rm e}^2\Lambda(T)},
\label{cooling}
\end{equation}
where $\mu m_{\rm p}$ is the mean molecular weight, $n_{\rm e}(r)$ is
the electron number density at the radius $r$, and $k$ is the Boltzmann constant.
Using the zero-metallicity and solar-metallicity cooling functions given by
Sutherland \& Dopita (1991), the cooling efficiency depending on
the metallicity of the gas are calculated by interpolation and extrapolation.

When a halo mass becomes more than double of the mass at the forming time,
the diffuse hot gas contained in the halo is reheated by shock to the virial
temperature of the halo \citep{som99}.  We refer this epoch as the
reheating time of the halo.
The cooling radius, $r_{\rm cool}$, is defined as a radius at
which the cooling time-scale, $\tau_{\rm cool}$, equals to the
elapsed time from the last reheating time of the halo, $t$.
The hot gas that distributes between $r_{\rm cool}(t)$ and
$r_{\rm cool}(t+\Delta t)$ is cooled and added to the cold gas
reservoir of the galaxy during the time-step, $\Delta t$.

The star formation in disks is described by the following simple law,
\begin{equation}
\dot{M}_* = \frac{M_{\rm cold}}{\tau_*},
\label{starrate}
\end{equation}
\begin{equation}
\tau_* = \tau_*^0\left(\frac{\tau_{\rm dyn}}{\tau_{\rm dyn}^0}\right),
\label{startime}
\end{equation}
where eq.(\ref{starrate}) denotes the rate of stars newly formed,
$\tau_*^0$ is the star formation time-scale of the galaxies which forms
at $z = 0$, $\tau_{\rm dyn} \equiv \frac{r_{\rm halo}}{V_{\rm c}}$ is
the dynamical time-scale of the galaxies,
and $\tau_{\rm dyn}^0$ indicates the dynamical time-scale of a halo
which forms at $z = 0$.
Here, we assume that the star formation time-scale is proportional
to the dynamical time-scale.
We calculate $\tau_{\rm dyn}^0$ according to the virial theorem and
the spherical collapse model.
The star formation time-scale at $z = 0$, $\tau_*^0$, is a free parameter
and we set to 5 Gyr in this paper.
We have confirmed that this parameter does not affect strongly on the morphology
of the cluster galaxies.

The feedback process by supernovae of massive stars has many uncertainties
actually, therefore we adopt a simple description \citep{col94},
\begin{equation}
\Delta M_{\rm reheat} =
\left(\frac{V_{\rm c}}{V_{\rm hot}}\right)^{-\alpha_{\rm hot}}
\dot{M}_* \Delta t \equiv \beta \dot{M}_* \Delta t,
\label{heat}
\end{equation}
where $V_{\rm hot}$ and $\alpha_{\rm hot}$ are free parameters. When adopting
$\alpha_{\rm hot} = 2$, eq.(\ref{heat}) corresponds to the formula
in Kauffmann et al. (1993), and Cole et al. (1994) used
$\alpha_{\rm hot} = 5.5$ in their fiducial model.

When two or more halos merge together, we identify a galaxy contained
in the largest progenitor as the central galaxy of the new common halo.
Other galaxies are identified as satellites.
These satellites merge with the central galaxy when the elapsed time
from the last reheating time exceeds the dynamical friction time-scale
\citep{bin87},
\begin{equation}
\tau_{\rm mrg} = \frac{1.17r_{\rm halo}^2V_{\rm c}}{\ln\Lambda GM_{\rm sat}},
\end{equation}
where $r_{\rm halo}$ and $V_{\rm c}$ are the radius and circular velocity
of the new common halo, respectively,
$M_{\rm sat}$ is the total mass of a halo to which the satellite belonged
as a central galaxy or the mass of stars and cold gas in the case that
the satellite was a halo-stripped galaxy, and $\ln\Lambda$ is
the Coulomb logarithm, which is approximated
as $\simeq \ln(1+M_{\rm halo}^2/M_{\rm sat}^2)$ \citep{som99}.
When the mass growth of the common halo satisfies the reheating condition,
the elapsed time from the last reheating time
is recalculated because orbits of satellites may be violently disturbed
in such a case.

When a satellite galaxy merges with a central galaxy and the mass ratio
of the galaxy with smaller mass to that with larger mass is larger than
$f_{\rm bulge}$, all stars of the satellite and disk stars of the central
galaxy are incorporated with the bulge of the central galaxy.
Then cold gases of both galaxies turns to stars in the bulge by {\it starburst}.
We  adopt $f_{\rm bulge} = 0.3$ for SCDM according to the numerical
simulations \citep{bar96}.
Since early formation of the cluster prevent galaxy mergers in OCDM,
we use $f_{\rm bulge} = 0.2$ for OCDM in order to reproduce the observed E+S0
fraction roughly (see \S 3).

\begin{table}[bth]
\begin{center}
\caption{Feedback parameters for models}
\begin{tabular}[htb]{ccc}
\tableline\tableline
Model & $V_{\rm hot}$ (km/s) & $\alpha_{\rm hot}$ \\
\tableline
A &280 &2 \\
B &140 &2 \\
C &200 &5.5 \\ \hline
\tableline
\label{model}
\end{tabular}
\end{center}
\end{table}

Chemical evolution is treated in almost the same way as described in Kauffmann
\& Charlot (1998).
The instantaneous recycling approximation is adopted.
The amount of metals ejected from supernovae is characterized by $y$,
which is heavy element yield for each generation of stars.
The fraction $f$ of the ejecta is ejected directly into the hot gas, and the rest is incorporated with the cold gas.
We adopt $y = 2{\rm Z_{\odot}}$ and $f = 0.3$, which are the same values as the strong feedback model in Kauffmann \& Charlot (1999).
The gas fraction returned by evolved stars, $R$, is 0.25 in this paper.
Simultaneously, the supernovae heat up the surrounding cold gas,
then metals contained in the cold gas are also returned to the hot gas.
The chemical evolution mainly affects on the colors of the galaxies.
Therefore, the value of the yield is not important in this paper.
The colors of the cluster galaxies will be discussed in our next paper.

In order to compare our results with observations directly,
stellar population synthesis model must be considered. We use the model
by Kodama \& Arimoto (1997) with the Salpeter's IMF having a slope of 1.35 and
mass range between $0.1{\rm M_{\odot}}$ and $60{\rm M_{\odot}}$.
The range of stellar metallicity $Z_*$ of simple stellar populations is
$0.0001 \sim 0.05$.
The luminosity of disks and bulges is given by this model in each band
according to the metalicities and ages of stars.
\begin{figure}[bht]
\rotatebox{90}{
\includegraphics[width=4.5cm,clip]{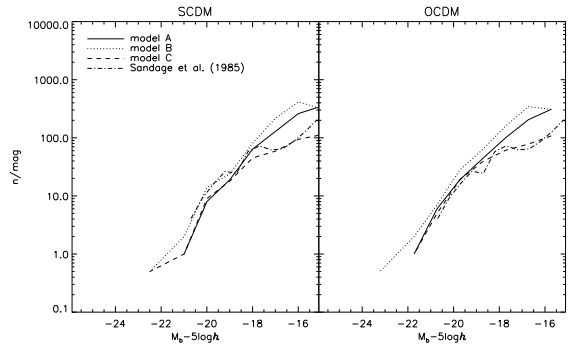}
}
\caption{
Luminosity functions of cluster galaxies.
The left and right panels show those in SCDM and OCDM, respectively, and
the thick solid, dotted, and dashed lines indicate those of the
model A, B, and C, respectively.
The thin dot-dashed line is the luminosity function of the Virgo
cluster galaxies by Sandage et al. (1985).}
\label{lumi}
\end{figure}

\subsection{Identification of morphology}
Morphology of each galaxy is determined by the $B$-band bulge-to-disk
luminosity ratio($B/D$).
In this paper, galaxies with $B/D \geq 1.52$ are identified
as ellipticals, $0.68 \leq B/D < 1.52$ as S0s, and $B/D < 0.68$ as spirals,
according to the results of Simien \& de Vaucouleurs (1986).
It has been shown that this method for classification reproduces observations
well by Kauffmann et al. (1993) and Baugh et al. (1996).

\subsection{Parameter Sets}
The feedback is key process which determines the feature of
galaxies (e.g. Kauffmann \& Charlot 1998). Therefore, we use
three type of feedback models in this paper (Table \ref{model}).
The model A is a strong feedback model and B is a normal feedback model.
In these two models $\alpha_{\rm hot}$ is set to 2.
In the model C, we adopt $\alpha_{\rm hot} = 5.5$ which is used in the fiducial
model of Cole et al. (1994), while $V_{\rm hot}$ is equal to 200 km s$^{-1}$
that is large compared to them.

In Fig. \ref{lumi}, we show the $B$-band luminosity functions of model
galaxies. The thick solid, dotted, and dashed lines indicate the luminosity
functions given by the models A, B, and C, respectively, and the thin dot-dashed
line indicates the observed luminosity function of the Virgo cluster
\citep{san85}.
The model C reproduces the observed luminosity function well by
flattening the faint-end slope of the luminosity function.  This is
caused by strong feedback to galaxies with low circular velocities
compared to the models A and B due to $\alpha_{\rm hot}=5.5$.
\begin{figure}[bht]
\rotatebox{90}{
\includegraphics[width=4.5cm,clip]{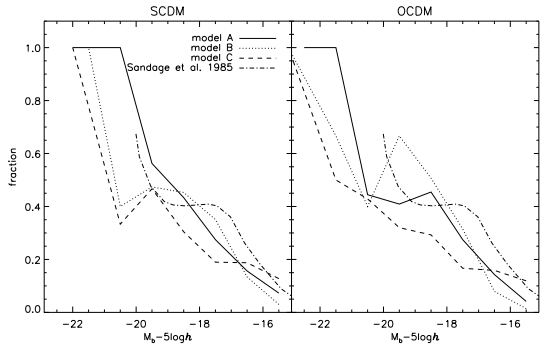}
}
\caption{
E + S0 fractions of cluster galaxies as a function of absolute $B$ magnitude
for SCDM (left panel) and OCDM (right panel).
The thick solid, dotted, and dashed lines indicate the model A, B, and C,
respectively.
The thin dash-dotted line shows the observational result for the Virgo cluster
taken from Sandage et al. (1985)
}
\label{mfrac}
\end{figure}

\section{Morphology of the Cluster Galaxies}
\subsection{Morphological Fraction}
In Fig. \ref{mfrac}, the E+S0 fractions of our cluster galaxies
are shown as a function of absolute B-band magnitude.
The observed curve for the Virgo cluster galaxies is taken from
Sandage et al. (1985).
All our models show the observational trend. It is, thus, found that
the merger-induced bulge formation naturally increases the fraction of
bulge-dominated galaxies toward a bright end.
In the model C, the clusters do not have early type galaxies sufficiently.
This is probably caused by the lack of S0 galaxies as we show below.

\subsection{Morphology-Density Relation}
\begin{figure}[hbt]
\begin{center}
\includegraphics[width=7.0cm,clip]{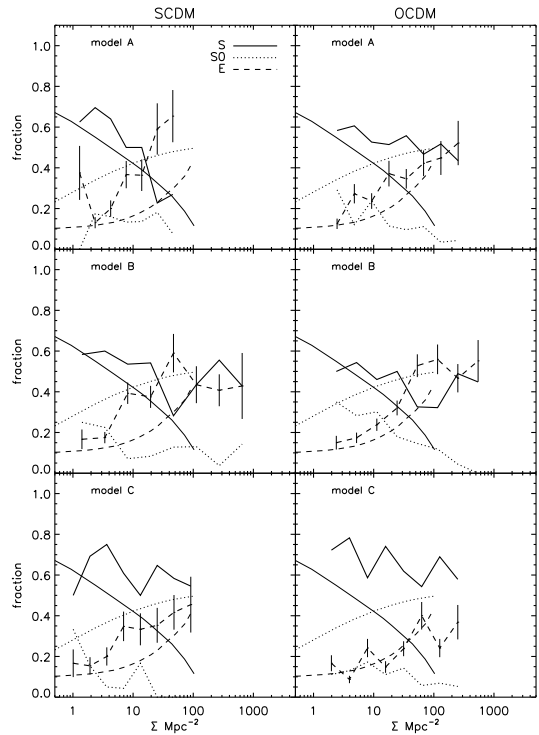}
\end{center}
\caption{
Morphology-density relations for cluster galaxies. The left and right panels
show the results for SCDM and OCDM, respectively.
The thick lines are the morphological fraction of 
the thin lines are those of the observed cluster galaxies by Dressler (1980).
S, E, and S0 fractions are represented by the solid, dotted, and dashed lines,
respectively.
We show the error bars for the elliptical fractions, which are 1$\sigma$
Poissonian uncertainties estimated from the number of galaxies in each bin.}
\label{md}
\end{figure}

The MDRs for the simulated clusters are represented in Fig. \ref{md}
using the same luminosity cutoff and definition of the local
projected density as Dressler (1980),
i.e. the local projected density is defined by nearest 10 neighbours
having the luminostiy $M_{V} - 5\log h < -18.9$.
The projected density is calculated in $x$--$y$, $y$--$z$, and $z$--$x$
projections for each model.
The morphological fractions of our models and Dressler's (1980) are
represented by thick and thin lines, respectively.
We attach the 1$\sigma$ error bars to the elliptical fractions according to
the number of the galaxies in each density bin.

In all models, the E fractions increase toward the high-density regions,
and this is consistent with the observed trend.
The S0 fractions, however, are much smaller than the observed
fraction and do not represent the observed trend of the moderate
increase toward the high-density regions.
We then examine whether such deficiency of S0 galaxies within the
simulated clusters can be solved by adjusting the $f_{\rm bulge}$
and the $B/D$ classification in the next subsection.

\begin{figure}[thb]
\includegraphics[width=7cm,clip]{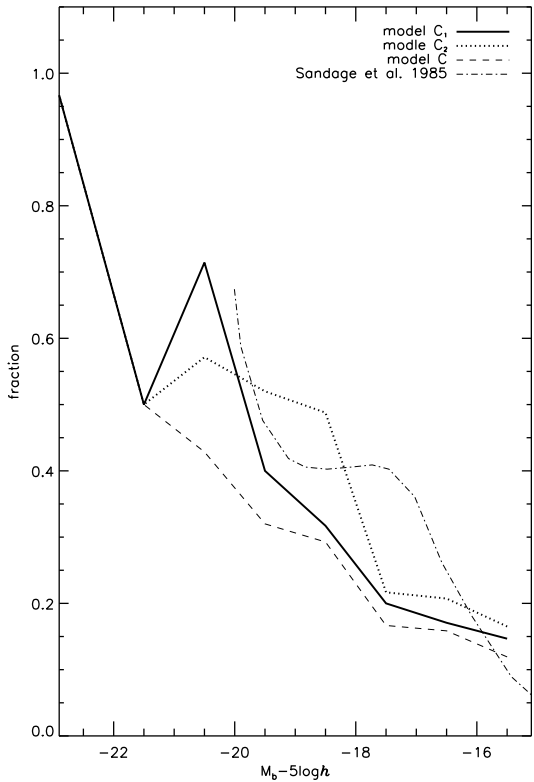}
\caption{
Same as Fig. \ref{mfrac}, but for the model C$_1$ (thick solid)
and C$_2$ (thick dotted).
For comparison, the fraction in model C and the fraction of the Virgo clusters
are represented by the thin dashed line and the thin dash-dotted lines,
respectively.}
\label{mfrac2}
\end{figure}

\subsection{Dependence on Model Parameters}
Since there is considerable scatter in the relationship between
the bulge-to-disk ratio and Hubble T-type (Baugh et al. 1996),
we should examine the case in which we adopt different values
of $B/D$ for the classification of the galaxies.
We also investigate how the parameter, $f_{\rm bulge}$, affects
on the MDRs.
\begin{figure}[bht]
\begin{center}
\includegraphics[width=7.0cm,clip]{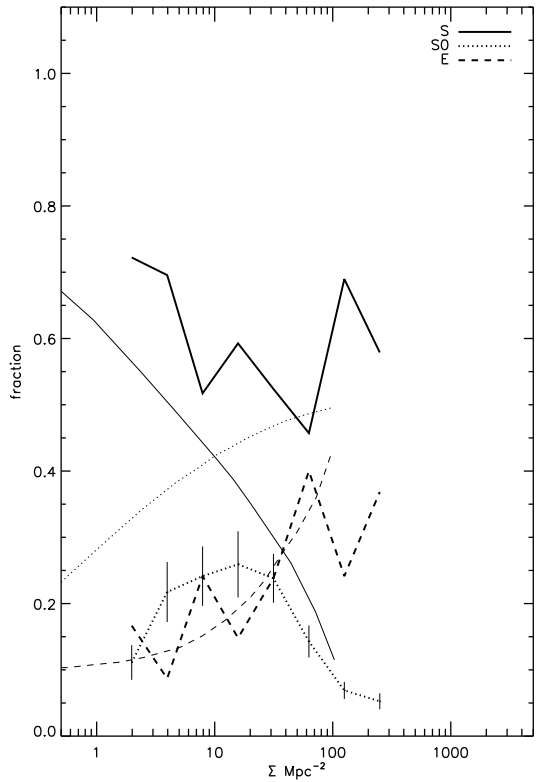}
\end{center}
\caption{
The MDR for the cluster in the model C$_1$.
We show the $1 \sigma$ error bars for the S0 fraction.
}
\label{md2}
\end{figure}

To study the influence of the choice of these parameters,
we use the model C in OCDM.
By this model, the luminosity function and the distribution of
the elliptical galaxies are in good agreement with the observations,
the E+S0 fraction, however, is insufficient.

First, we adapt the new $B/D$ classification, that is,
galaxies with $B/D \geq 1.52$ are identified as
ellipticals, $0.25 \leq B/D < 1.52$ as S0s,
and $B/D < 0.25$ as spirals.
We call this model the model C$_1$.
In Fig. \ref{mfrac2}, we plot the E+S0 fraction of this model
(solid line).
The fraction is slightly increased, however, still below the observation.
The MDR of this model is presented in Fig. \ref{md2}.
Although the S0 fraction is increased by adopting wider $B/D$ range for S0s,
the fraction is too small, especially in the high-density regions.
Koopmann \& Kenney (1998) indicated that galaxies with low central
concentration (i.e. low $B/D$), with which they should be classified
into spiral  galaxies such as Sa galaxies in field environment,
were often identified as S0 galaxies due to their low star formation rates.
They suggested that misleading classification of the low central-concentration
galaxies as S0s might account for a part of domination of S0 galaxies
in cluster environment.
Our result shows that the distribution of the S0s is not
reproduced by the merger-driven bulge formation
even when we identify the galaxies with low $B/D$ as S0s.
\begin{figure}[bht]
\begin{center}
\rotatebox{90}{
\includegraphics[width=4.5cm,clip]{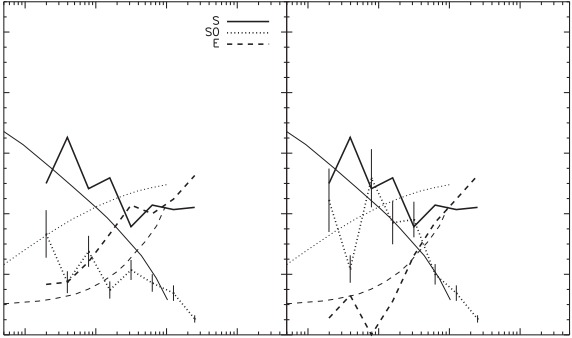}
}
\end{center}
\caption{
The MDRs for the cluster in the model C$_2$.
In the left panel, we show the MDR adopting the morphological classification
mentioned in \S 2.3.
In the right panel, we represnt the MDR in the case that we identify the
galaxy with $0.68 \le B/D < 4$ as S0s.
We show the $1 \sigma$ error bars for the S0 fractions.
}
\label{md3}
\end{figure}

Next, we set $f_{\rm bulge}$ to 0.1 in order to increase the
E+S0 fraction.
Note that this value of $f_{\rm bulge}$ is extremely low.
Indeed, a merger with the mass ratio of 0.1 should be considered
as a minor merger \citep{wak96}.
We call this model the model C$_2$.
The E+S0 fraction of model C$_2$ with the $B/D$ classification
in \S2.3 is shown in Fig. \ref{mfrac2}.
The fraction is fairly increased than the model C and show better
agreement with the observation.
In the left panel of the Fig. \ref{md3}, we show the MDR of
this model. It is found that both of E and S0 fraction are increased,
the shortage of S0s is, however, still significant.
We then examine the case in which we change the threshold value
of the $B/D$ between E and S0 in order to increase the S0 fraction.
We identify the galaxies with $B/D \geq 4$ as Es, and then the galaxies
with $0.68 \leq B/D < 4$ as S0s.
The MDR in this case is indicated in the right panel of Fig. \ref{md3}.
The S0 fraction becomes comparable with the observation in the
low-density regions, whether it hardly changes in the high-density regions.
It means that the galaxies in very high-density regions mainly separate
into two types, i.e.  bulge dominated and disk dominated galaxies,
under the {\it merger hypothesis}.

\section{DISCUSSION}
In this paper, we represent a new method that combines high-resolution
$N$-body simulations with a semi-analytic galaxy formation model
for cluster  galaxies.
The high-resolution simulations enable us to identify and trace galactic
halos within clusters directly.
Therefore, we can obtain the three-dimensional distribution of the cluster
galaxies and also we can incorporate the dynamical processes, for example
merging and tidal stripping of the galactic halos in the clusters, into
our modeling of galaxy formation.
For the first study of the cluster galaxies using this method,
we have shown the MDRs for the simulated clusters of galaxies with
two cosmological models based on the merger-driven bulge formation scenario.

The model which reproduce the observed luminosity function well (model C) is
successful in explaining the observed elliptical fraction as a function of
the local density.
However, any models cannot reproduce the observed trend in
the distribution of the cluster S0s.
In all models, the S0 fractions are too small, especially in the high density
regions.

We also examine the case in which the S0 galaxies in the cluster environment
have the different $B/D$ range from the S0 galaxies in the field, because
Koopmann \& Kenney (1998) found that the galaxies with low central
concentration, which should be identified as spiral galaxies in the field,
were often classified into the S0 type in clusters.
We have confirmed that the S0 fraction are increased only in the low density
regions even if we adopt lower threshold value of $B/D$ between S and S0.

We then investigate how the parameter, $f_{\rm bulge}$, affects on the MDR.
When we choose smaller $f_{\rm bulge}$, the early-type fraction is increased.
In this case, however, the elliptical fraction is mainly increased, the shortage of S0s is then still significant.
Even if we classify the low $B/D$ ellipticals as S0s in the model with small
$f_{\rm bulge}$, this change increases the S0 fraction only in the low density
regions again.

Above results imply that, under the merger-driven bulge formation
model, the galaxies separate into two types, i.e. almost pure bulge
and pure disk systems, in the high density regions independent on the
choice of the parameters and cosmologies.

The reason is probably considered as follows.
In the model adopted here, a S0 galaxy is  only formed by the disk formation
after the last major merger.
The efficiency of merging is, however, high before cluster formation,
and then it rapidly decreases due to
the large internal velocity dispersions of the clusters and the reduction of
the size of tidally truncated halos as the mass of clusters grows.
The accretion onto the galactic halos is also prevented by the
strong tidal field after beginning of the cluster formation \citep{oka99}.
Hence, the cluster ellipticals formed by the mergers at high redshifts
hardly change their morphologies into S0s by minor mergers or gas accretion,
which form additional stellar disks.

Our results suggest that the {\it merger hypothesis} can give a good
explanation of the distribution of cluster ellipticals.
On the other hand, to reproduce the observed S0 fraction, the environmental
effects that change the morphology of spiral galaxies into S0 galaxies
directly should be considered, for example
the ram-pressure stripping \citep{gun72, fn99},
the galaxy harassment (Moore et al. 1996, 1998), and the minor mergers
\citep{wak96}.

The morphological classification based on $B/D$ is the simplest way.
There are various ways of classification other than the $B/D$ classification.
For instance, Caon \& Einasto (1995) found another type of morphological
segregation of early-type galaxies in the Virgo cluster,
based on the configurations of their isophotes.
Comparing local projected densities, galaxies with boxy isophotes are
located in local density enhancements, while galaxies with disky isophotes
lie in regions of lower local densities, independently of their classification
as elliptical or lenticular.
Although, it is interesting to investigate what physical processes are
responsible for this segregation, more numerical studies are needed to
include such morphological classification in our model because we need to
understand the dynamical states of the internal structures of galaxies
to modeling it, and it is then left for further studies.

In the next paper, we will investigate whole properties of cluster galaxies
in detail, i.e. colors, metalicities, velocities, and so on,
using the method presented in this paper.
We will also discuss some environmental effects.


\acknowledgments

We are grateful to A. Habe, Y. Fujita, and T. Saito for
useful discussions and valuable insights.
Numerical computation in this work was carried out on the HP Exemplar
at the Yukawa Institute Computer Facility and on the SGI Origin 2000
at the Div. of Physics, Grad. School of Science, Hokkaido Univ.

\clearpage


\begin{thebibliography}{}
\bibitem[Barnes 1996]{bar96} Barnes, J. E.1996, in ASP Conference Series 92,
	Formation of the Galactic Halo---Inside and Out,
	ed. H. L. Morrison \& A. Salajedini (San Francisco: ASP), 415
\bibitem[Baugh et al. 1996]{bau96} Baugh, C. M., Cole, S.,
	\& Frenk, C. S. 1996, \mnras, 283, 1361
\bibitem[Binney \& Tremaine 1987]{bin87} Binney, J., \& Tremaine, S. 1987,
	Galactic Dynamics, (Princeton: Princeton Univ. Press)
\bibitem[Benson et al. 2000]{ben00} Benson, A. J., Cole, S., Frenk, C. S.,
	Baugh, C. M., \& Lacey, C. G. 2000, \mnras, 311, 793
\bibitem[Bond et al. 1991]{bon91} Bond, J. R., Cole, S., Efstathiou, G., \&
	Kaiser, N. 1991, \apj, 379, 440
\bibitem[Bower 1991]{bow91} Bower, R. G. 1991, \mnras, 248, 332
\bibitem[Caon \& Einasto 1995]{cao95} Caon, N., \& Einasto, M. 1995,
	\mnras, 273, 913
\bibitem[Cole et al. 1994]{col94} Cole, S., Arag\'{o}n-Salamanca, A.,
	Frenk, C. S., Navarro, J. F., \& Zepf, S. E. 1994, \mnras, 271, 781
\bibitem[Couch et al. 1994]{cou94} Couch, W. J., Ellis R. S., Sharples, R. M.,
	\& Smail, I. 1994, \apj, 430, 121
\bibitem[Dressler 1980]{dre80} Dressler A. 1980, \apj, 236, 351
\bibitem[Dressler et al. 1997]{dre97} Dressler A. et al. 1997, \apj, 490, 577
\bibitem[Evrard et al. 1994]{evr94} Evrard, A. E., Summers, F. J.,
	\& Davis, M. 1994, \apj, 422, 11
\bibitem[Fujita \& Nagashima 1999]{fn99} Fujita, Y., \& Nagashima, M., 1999, \apj, 516, 619
\bibitem[Gunn \& Gott 1972]{gun72} Gunn, J. E., \& Gott, J. R. 1972, \apj, 176, 1
\bibitem[Katz et al. 1992]{kat92} Katz, N., Hernquist, L., \& Weinberg, D. H. 1992, \apj, 399, L109
\bibitem[Kauffmann 1996]{kau96} Kauffmann, G. 1996, \mnras, 281, 487
\bibitem[Kauffmann \& Charlot 1998]{kc} Kauffmann, G., \& Charlot, S. 1998, \mnras, 294, 705
\bibitem[Kauffmann et al. 1999]{kau99} Kauffmann, G., Colberg, J. M., Diaferio, A., \&
	White, S. D. M. 1999, \mnras, 397, 529
\bibitem[Kauffmann et al. 1993]{kau93} Kauffmann, G., White, S. D. M.,
	\& Guiderdoni, B. 1993, \mnras, 264, 201
\bibitem[Kodama \& Arimoto 1997]{kod97} Kodama, T., \& Arimoto, N. 1997,
	\aap, 320, 41
\bibitem[Koopmann \& Kenney 1998]{koo98} Koopmann, R. A., \& Kenney, J. D. P.
	1998, \apj, 497, L75
\bibitem[Lacey \& Cole 1993]{lac93} Lacey, C., \& Cole, S. 1993, \mnras, 262, 627
\bibitem[Moore et al. 1996]{moo96} Moore, B., Katz, N., Lake, G., Dressler, A.,
	\& Oemler, A., 1996, \nat, 379, 613
\bibitem[Moore et al. 1998]{moo98} Moore, G., Lake, G., \& Katz, N. 1998, \apj, 495, 139
%
%
\bibitem[Okamoto \& Habe 1999]{oka99} Okamoto, T., \& Habe, A. 1999, \apj, 516, 591
\bibitem[Okamoto \& Habe 2000]{oka00} Okamoto, T., \& Habe, A. 2000, \pasj, 52,
457
\bibitem[Press \& Schechter 1972]{pre72} Press, W. H., \& Schechter, P. L. 1974, \apj, 187, 425
%
%
\bibitem[Roukema et al. 1997]{rou97} Roukema, B. F., Peterson, B. A., Quinn, P. J., \&
	Rocca-Volmerrange, B. 1997, \mnras, 292, 835
\bibitem[Salpeter 1955]{sal55} Salpeter, E. E. 1955, \apj, 121, 161
\bibitem[Sandage et al. 1985]{san85} Sandage, A., Binggeli, B., \& Tammann, G. A. 1985, \aj, 90, 1759
\bibitem[Simien \& de Vaucouleurs 198)]{sim86} Simien, F., \& de Vaucouleurs, G. 1986, \apj, 302, 564.
\bibitem[Somerville \& Primack 1999]{som99} Somerville, R. S., \& Primack, J. R. 1999, \mnras, 310, 1087
\bibitem[Summers et al. 1995]{sum95} Summers, F. J., Davis, M., \& Evrard, A. 1995, \apj, 454, 1
\bibitem[Sutherland \& Dopita 1993]{sut93} Sutherland, R., \& Dopita, M. A. 1993, \apjs, 88, 253
\bibitem[Toomre 1977]{too77} Toomre, A., 1977, in
	The Evolution of Galaxies and Stellar Populations,
	ed. B. M. Tinsley \& R. B. Larson (New Haven: Yale Univ. Press), 401
\bibitem[Walker et al. 1996]{wak96} Walker, I. R., Mihos, J. C., \& Hernquist, L. 1996, \apj, 460, 121
\end{thebibliography}
\end{document}